\documentstyle[a4,12pt,epsf,amssymb]{article}
\begin{document}
\textheight 23cm
\topmargin -0.5cm
\textwidth 16cm
\oddsidemargin 0.65cm
\evensidemargin 0.65cm
\setlength{\parskip}{0.45cm}
\setlength{\baselineskip}{0.75cm}
%
\begin{titlepage}
\begin{flushright}
CERN-TH/97-270 \\
DO-TH 97/21 \\
TPR-97-18 \\
October 1997 \\ 
\end{flushright}
\begin{center}
\renewcommand{\thefootnote}{\fnsymbol{footnote}}
\setcounter{footnote}{1}
\vspace*{1.cm}
{\LARGE
{\bf Soffer's inequality and the transversely  }  

\vspace{-0.2cm}
{\bf polarized Drell-Yan process at} 

\vspace{-0.2cm}
{\bf next-to-leading order}  } 

\vspace{0.1cm}

\vspace*{1cm}
{\Large O.~Martin$^a$, A.~Sch\"afer$^a$, 
M.~Stratmann$^{b,\,}$\footnote{Address after October 1$^{\rm st}$, 1997: 
Department of Physics, University of Durham, Durham, DH1 3LE, England}, 
W.~Vogelsang$^c$}
\vspace{1cm}
\linebreak
{a) Institut f\"ur Theoretische Physik, Universit\"at Regensburg, 
D-93040~Regensburg, Germany
\linebreak
b) Institut f\"ur Physik, Universit\"at Dortmund, D-44221~Dortmund, Germany
\linebreak
c) Theory Division, CERN, CH-1211~Geneva~23, Switzerland}
\vspace*{2cm}
\linebreak
\renewcommand{\thefootnote}{\fnsymbol{roman}}
\setcounter{footnote}{0}
{\bf Abstract}

\vspace{-0.3cm}
\end{center}
We check numerically 
if Soffer's inequality for quark distributions is preserved by
next-to-leading order QCD evolution. Assuming that the inequality is 
saturated at a low hadronic scale we estimate the maximal transverse double 
spin asymmetry for Drell-Yan muon pair production to next-to-leading order 
accuracy.
\end{titlepage}
\section{Introduction}
The transversity distribution $\delta q(x,Q^2)$ is the only completely 
unknown twist-2 parton distribution function of the nucleon. In a 
transversely polarized nucleon it counts the number of quarks  
with spin parallel to the nucleon spin minus the number of quarks with 
antialigned spin \cite{raso}. In field theory the transversity 
distribution is defined by the expectation value of a chiral-odd 
operator between nucleon states which is the reason why it is not 
experimentally accessible in inclusive deep inelastic 
lepton-nucleon scattering (DIS) \cite{jaji,arme}. 
The most promising hard process allowed by this 
chirality selection rule seems to be Drell-Yan dimuon production,  
and exactly this reaction will be utilized for attempting a first measurement
of $\delta q(x,Q^2)$ at RHIC \cite{rhic}. 
What actually will be measured is not 
the transversity
distribution itself, but the transverse double spin asymmetry 
$A_{TT}=d\delta \sigma/d\sigma$ where the polarized and 
unpolarized hadronic cross sections 
are defined as
\begin{equation}
d\delta\sigma \equiv \frac{1}{2} \left( 
d\sigma^{\uparrow\uparrow}-d\sigma^{\uparrow\downarrow}\right)\quad,
\quad\quad
d\sigma \equiv \frac{1}{2} \left( 
d\sigma^{\uparrow\uparrow}+d\sigma^{\uparrow\downarrow}\right)\quad.
\end{equation}
In perturbative QCD (pQCD) $A_{TT}$ can be expressed in terms of unpolarized
parton distributions, the yet unknown transversity distributions and the
relevant partonic cross sections. Although the latter have been known to 
next-to-leading order (NLO) accuracy in the strong coupling for several 
years by now \cite{fupe,vowe,coka}, consistent NLO calculations
were not possible because of the lack of the two-loop transversity splitting 
functions. This situation changed only very recently \cite{voge,haka,kumi}, 
allowing for the first time for a consistent calculation of pQCD 
corrections to the transverse double spin asymmetry for the Drell-Yan
process.

The unpolarized, longitudinally and transversely polarized quark 
distributions ($q$, $\Delta q$, $\delta q$) of
the nucleon are expected to obey the rather interesting relation
\begin{equation}
2 |\delta q(x)| \leq q(x) + \Delta q(x) \label{soffer}
\end{equation}
derived by Soffer \cite{soff}. 
It has been recently clarified that Soffer's inequation is 
preserved by leading order (LO) QCD evolution, i.e.\ if (\ref{soffer}) 
is valid at some scale $Q_0$ it will also be valid at $Q>Q_0$ \cite{baro}.
To NLO the situation is not as simple. The parton distributions are now 
subject to the choice of the factorization scheme which one may fix
independently for $q$, $\Delta q$ and $\delta q$. One can therefore always 
find ``sufficiently incompatible'' schemes in which a violation of 
(\ref{soffer}) occurs. However, in \cite{voge}
it was shown with analytical methods that the inequation for valence 
densities is preserved by NLO QCD evolution in a certain 
``Drell-Yan scheme'' in which the NLO cross sections for dimuon production 
maintain their LO forms, and also in the $\overline{\mbox{MS}}$ 
scheme. An analytical check of the sea part is difficult since 
the singlet mixing between quarks and gluons has to be taken into account for
the unpolarized and longitudinally polarized quantities on the 
right-hand-side (r.h.s.) of (\ref{soffer}). In Section~\ref{evol} of 
this article  we shall show numerically that Soffer's inequation 
for sea quarks is also preserved under NLO evolution.

Estimates of $A_{TT}$ suffer of course from the fact that no 
experimental information on the transversity distribution is 
available at the moment. Therefore one has to rely on ans\"atze or 
model calculations of $\delta q(x,Q^2)$ at some reference scale $Q_0$  
\cite{masc}. For example, a popular assumption is $\delta q(x,Q^2_0)=
\Delta q(x,Q_0^2)$ which, however, is in general incompatible with Soffer's 
inequality (\ref{soffer}), in particular in a situation in which 
$\Delta q (x,Q_0^2) \approx - q(x,Q_0^2)$. Our aim in 
Sections~\ref{bounds} and \ref{results} will be to estimate within LO and 
NLO an upper bound on the transverse double spin asymmetry for the 
Drell-Yan process. To do so, we will first of all assume validity of 
Soffer's inequality which seems reasonable and is corroborated by our 
finding of Section~\ref{evol} that the NLO evolution to $Q^2>Q_0^2$ 
preserves the inequation once it is satisfied at the input scale. 
The maximal asymmetry $A_{TT}$ can then be estimated by further assuming
{\em saturation} of the Soffer bound (\ref{soffer}). The result obtained for
$A_{TT}$ under this assumption obviously strongly depends on the 
value chosen for $Q_0$. If $Q_0$ is taken to be large, i.e.\ of the order 
of the invariant mass $M$ of the lepton pair which sets the typical hard 
scale for the Drell-Yan process, the largest possible values for $A_{TT}$ 
will be reached. However, for several reasons it does not seem convincing to
assume saturation of (\ref{soffer}) by the input distributions employing such
a high $Q_0^2\sim M^2$: Firstly, evolving backwards to 
$Q^2 < Q_0^2$ -- which should
be a completely legitimate procedure if $Q_0$ is not small -- will under such 
circumstances immediately yield a violation of Soffer's inequality. 
Secondly, the r.h.s. of (\ref{soffer}) will almost certainly 
lead to an overestimation for the $\delta q$ if saturation is assumed
at a large $Q_0$. For instance, for sea quarks the first moment 
($x$-integral) of the r.h.s. in (\ref{soffer}) will diverge, 
which is not expected for the integral over $\delta \bar{q}$ at any
$Q^2$. Therefore, to obtain a realistic estimate for an upper bound on 
$A_{TT}$ by assuming saturation of Soffer's inequation, two requirements 
have to be met: (i) The saturation should be adopted only at a rather low 
``hadronic'' input scale where (ii) the integral over the r.h.s. of
(\ref{soffer}) is finite. Both demands are automatically fulfilled if we
choose the unpolarized and longitudinally polarized input parton 
distributions of the radiative parton model analyses \cite{glre1,glre2,glre3} 
and set\footnote{The possibility of choosing a different sign in front of the 
r.h.s. of (\ref{soffer1}) will be discussed later.}
\begin{equation}
2 \delta q(x,Q_0^2) = q(x,Q_0^2) + \Delta q(x,Q_0^2) \label{soffer1} \; ,
\end{equation}
where $Q_0$ now is identified with the input scale $\mu \sim {\cal O} (0.6$ 
GeV) of the radiative parton model \cite{glre1} and is considered the 
smallest scale from which perturbative evolution can be performed, such that 
no backward evolution from $\mu$ makes sense. While we are aware that our 
approach with its rather small $Q_0$ may lead to an underestimation of the 
maximally possible $A_{TT}$, we still believe our results to be built on
a firm basis, given the large phenomenological success \cite{glre2,glre3} 
of the radiative parton model for the $q$, $\Delta q$. In any case, our 
results for $A_{TT}$ under the assumption of (\ref{soffer1}) are the largest
the radiative parton model can predict and will provide a useful target 
for future experiments. We emphasize that our NLO results presented in 
Section~\ref{results} are the first ones to be obtained to true and 
consistent NLO accuracy. Section~\ref{results} will also provide a 
discussion of other possible uncertainties of our results. 

In Section~\ref{conclusions} we will present our conclusions.

\section{Preservation of Soffer's inequation by NLO evolution\label{evol}}
Unlike the case of the more familiar unpolarized and longitudinally polarized
densities, all transversity distributions obey simple non-singlet type 
evolution equations because there is no corresponding gluonic quantity  
due to angular momentum conservation \cite{jama,arme}. Introducing
\begin{equation} \label{pm}
\delta q_{\pm}(x,Q^2)\equiv \delta q(x,Q^2) \pm \delta \bar q(x,Q^2)
\end{equation}
and Mellin moments $\delta q^n_{\pm}(Q^2)\equiv\int_0^1 dx x^{n-1} 
\delta q_{\pm}(x,Q^2)$ the evolution equations are given by \cite{fupe}
\begin{equation}
Q^2\frac{d}{dQ^2} \delta q_{\pm}^n(Q^2) = \delta P^n_{qq,\pm}(\alpha_s(Q^2))
\delta q_{\pm}^n(Q^2)\quad . \label{eveq}
\end{equation}
The Mellin moments of the transverse splitting functions $\delta 
P^n_{qq,\pm}$ are taken to have the following perturbative expansion
\begin{equation}
\delta P_{qq,\pm}^n (\alpha_s) = \left( \frac{\alpha_s}{2\pi} \right)
\delta P_{qq}^{(0),n} + \left(\frac{\alpha_s}{2\pi}\right)^2
\delta P_{qq,\pm}^{(1),n} + \dots \quad,
\end{equation}
i.e.\ both are equal to LO. We use the following NLO expression for the 
strong coupling constant,
\begin{equation}
\frac{\alpha_s(Q^2)}{2\pi} = \frac{2}{\beta_0 \ln Q^2/\Lambda^2}
\left( 1 - \frac{\beta_1}{\beta_0^2} 
\frac{\ln\ln Q^2/\Lambda^2}{\ln Q^2/\Lambda^2} \right) \; ,
\end{equation}
where $\Lambda$ is the QCD scale parameter and $\beta_0=11-2n_f/3$, 
$\beta_1=102-38n_f/3$ with $n_f$ being the number of active flavors. 
The solution of (\ref{eveq}) is then simply given by \cite{fupe}
\begin{eqnarray}
\lefteqn{\delta q^n_{\pm}(Q^2) 
=  \left[ 1+ \frac{\alpha_s(Q^2_0)-\alpha_s(Q^2)}
{\pi\beta_0} \left( \delta P_{qq,\pm}^{(1),n} - \frac{\beta_1}{2\beta_0}
\delta P_{qq}^{(0),n} \right) \right]
} \hspace{2cm} \nonumber
\\
& & \hspace{-0.6cm}
\times
\left( \frac{\alpha_s(Q^2)}{\alpha_s(Q_0^2)}\right)^{-2\delta P_{qq}^{(0),n}
/ \beta_0} \delta q^n_{\pm}(Q_0^2) \quad . \label{solution}
\end{eqnarray}
Needless to say that the LO evolutions are entailed in the above 
equations when we put the NLO quantities, $\delta P_{qq,\pm}^{(1),n}$,
$\beta_1$, to zero. 

Eq.~(\ref{solution}) can be very conveniently employed for a numerical 
calculation of the NLO evolution of the transversity distributions. As
discussed in the introduction, we will assume saturation of Soffer's
inequality at the input scale, see Eq.~(\ref{soffer1}). Our choice for the 
r.h.s. of (\ref{soffer1}) will then be the NLO $\overline{\mbox{MS}}$ 
radiative parton model inputs for $q(x,Q_0^2)$ of \cite{glre2} and for the 
longitudinally polarized $\Delta q(x,Q_0^2)$ of the ``standard'' scenario 
of \cite{glre3} at\footnote{Note that for the purpose of checking 
the preservation of Soffer's inequality by evolution the choice of the 
initial scale $Q_0^2$ is actually irrelevant.} $Q_0^2=\mu_{NLO}^2=0.34$ 
GeV$^2$. For simplicity we will slightly deviate from the actual $q(x,Q_0^2)$
of \cite{glre2} in so far as we will neglect the breaking of SU(2) in the 
input sea quark distributions originally present in this set. This seems
reasonable as SU(2)-symmetry was also assumed for the $\Delta \bar{q}
(x,Q_0^2)$ of \cite{glre3}, which in that case was due to the fact that
in the longitudinally polarized case there are no data yet that could
discriminate between $\Delta \bar{u}$ and $\Delta \bar{d}$. We therefore
prefer to assume $\delta \bar{u} (x,Q_0^2)=\delta \bar{d} (x,Q_0^2)$ also
for the transversity input. We will examine the possible effects of 
SU(2)-breaking later. The moments of the resulting input distributions 
$\delta q(x,Q_0^2)$ are easily taken, and 
the $\delta q^n_{\pm} (Q_0^2)$ are then evolved to higher scales $Q^2>Q_0^2$ 
with the help of (\ref{solution}). A standard inverse Mellin transformation 
finally gives the desired transversity distribution in $x$-space. 
In order to perform this inverse Mellin transformation, Eq.~(\ref{solution}) 
has to be analytically continued to complex $n$ \cite{glre1}. The evolutions
of the $q(x,Q_0^2)$ (neglecting the SU(2)-breaking) and the $\Delta
q(x,Q_0^2)$, which both involve the singlet mixing between quarks and gluons, 
proceed as explained in \cite{glre1,glre2,glre3}.

\begin{figure}
\epsfysize9cm
\hspace*{0.9cm}
\leavevmode\epsffile{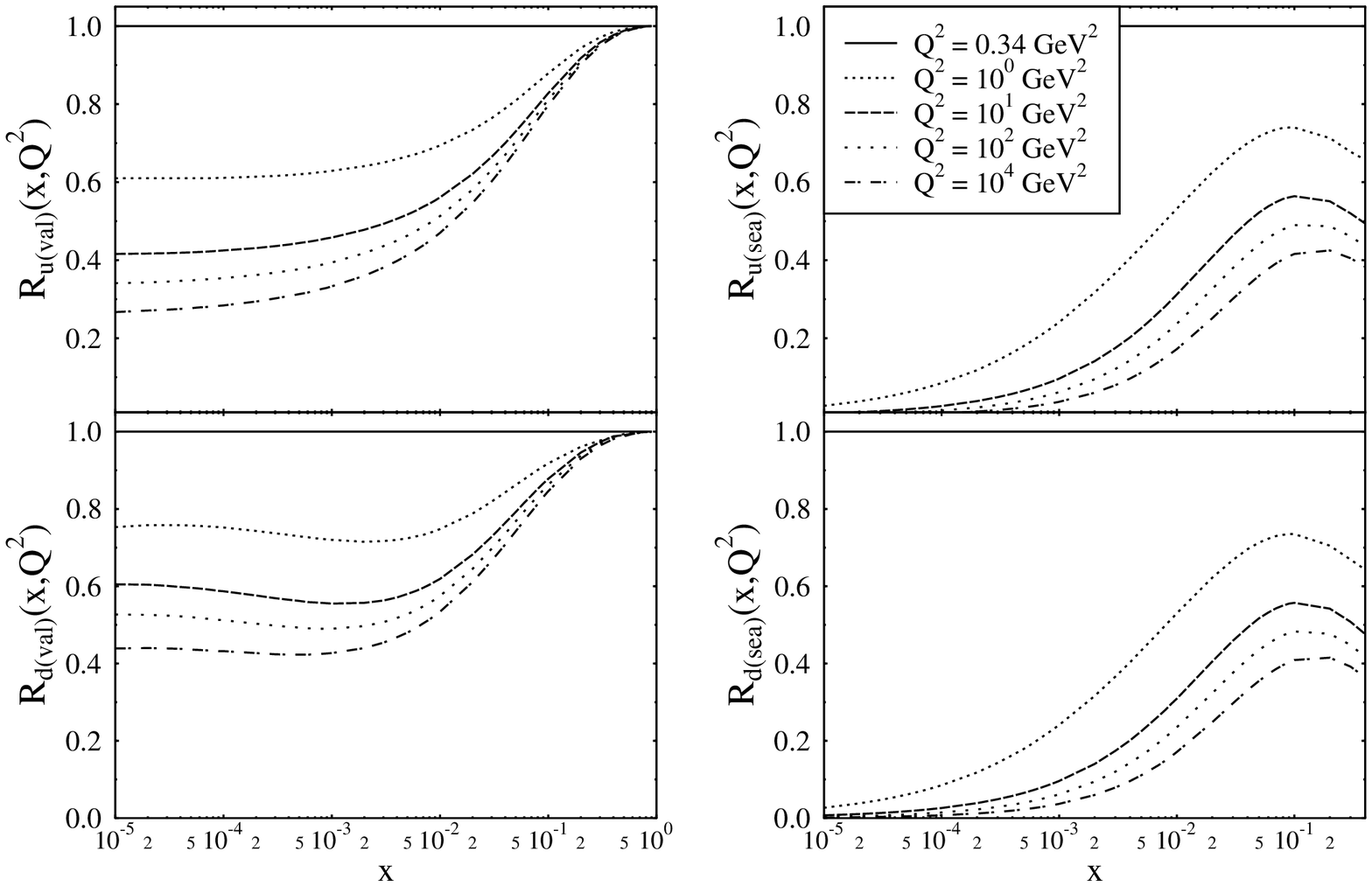}
\caption{\label{fig1}The ratio $R_q(x,Q^2)$ as defined in (\ref{req}) 
for $q=u_v,\bar{u},d_v,\bar{d}$ and several fixed values of $Q^2$.}
\end{figure}

In order to numerically check the preservation of (\ref{soffer}), 
Fig.~\ref{fig1} shows the ratio
\begin{equation}
R_q(x,Q^2) = \frac{2|\delta q(x,Q^2)|}{q(x,Q^2)+\Delta q(x,Q^2)}
\label{req}
\end{equation}
as a function of $x$ for several different $Q^2$ values for $q=u_v=u_-$, 
$\bar{u}=(u_+-u_-)/2$, $d_v=d_-$, $\bar{d}=(d_+-d_-)/2$ (cf. Eq.~(\ref{pm})).
If NLO evolution preserves Soffer's inequality then $R_q(x,Q^2)$ should not 
become larger than 1 for any $Q^2\geq Q_0^2$. As we already know from 
\cite{voge} this is the case for the two valence 
distributions. Fig.~\ref{fig1} confirms that Soffer's inequality is 
indeed also preserved for sea distributions. Furthermore, in Fig.~\ref{fig1} 
we see that evolution leads to a strong suppression of 
$R_q(x,Q^2)$ at small values of $x$, in particular for the sea quarks. 
This can be understood by the fact that $\delta P_{qq,\pm}(x)$ has a very 
mild behaviour for $x\rightarrow 0$ \cite{voge}, and by the well-known sharp 
small-$x$ rise of the unpolarized sea distributions in the denominator 
of $R_q$ due to $Q^2$-evolution. We note that our numerical results for 
the sea quarks became somewhat unstable at large $x$, probably caused
by the fact that the sea distributions are obtained here as differences
of two much larger quantities. 

As is obvious from (\ref{soffer}), Soffer's inequality  only restricts the 
absolute value of the transversity distribution. Therefore, we are free to 
choose a different sign in front of the r.h.s.\ of (\ref{soffer1}) and have 
to check the results for the two distinct possibilities 
$\delta q_v(x,Q_0^2)>0$,~$\delta \bar{q}(x,Q_0^2)>0$ 
and $\delta q_v(x,Q_0^2)>0$,~$\delta \bar{q}(x,Q_0^2)<0$. 
Our results do not noticeably depend on the actual choice.

\begin{figure}
\epsfysize9cm
\hspace*{3cm}
\leavevmode\epsffile{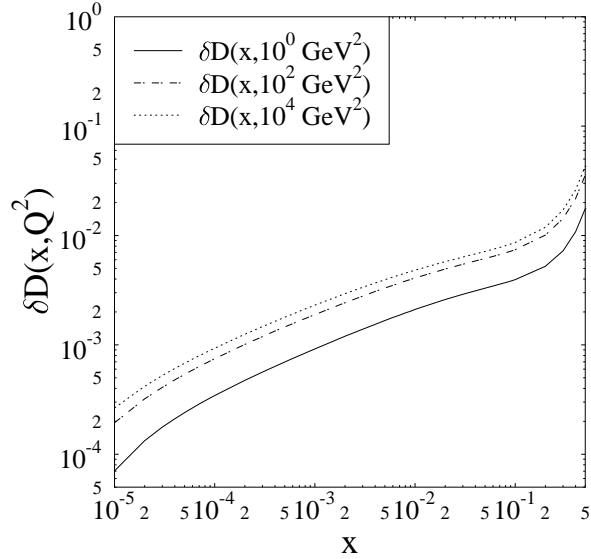}
\caption{\label{fig2}The dynamical SU(2)-breaking in the NLO transversity 
densities expressed by the ratio $\delta D(x,Q^2)$ as defined in 
(\ref{xxx}) for several fixed values of $Q^2$.}
\end{figure}

As we have neglected any possible SU(2)-breaking in all the sea input
distributions $\bar{q} (x,Q_0^2)$, $\Delta \bar{q} (x,Q_0^2)$,
$\delta \bar{q} (x,Q_0^2)$, any difference between the curves for 
$R_{\bar{u}}$, $R_{\bar{d}}$ can necessarily only result from the
dynamical breaking of SU(2) first induced by NLO evolution. The occurrence
of a small breaking from this source is well-known from the unpolarized
\cite{ross} and longitudinally polarized \cite{svw} cases. For the 
transversity densities it is given by
\begin{eqnarray}
\nonumber
2 \left( \delta \bar{u}-\delta \bar{d} \right)^n (Q^2)
&=& \frac{\left(\alpha_s (Q^2)-\alpha_s (Q_0^2)\right)}{\pi\beta_0}
\left(  \delta P_{qq,-}^{(1),n} - \delta P_{qq,+}^{(1),n} \right)\\
&&\times
\left( \frac{\alpha_s (Q^2)}{\alpha_s (Q_0^2)}
\right)^{-2\delta P_{qq}^{(0),n}/ \beta_0} \left( \delta u_v -
\delta d_v \right)^n (Q_0^2)  \:\:\: .
\end{eqnarray}
Fig.~\ref{fig2} displays the resulting effect via the ratio
\begin{equation} \label{xxx}
\delta D(x,Q^2)=\frac{\delta \bar{u}(x,Q^2) -\delta \bar{d} (x,Q^2)}
{\delta \bar{u}(x,Q^2) + \delta \bar{d} (x,Q^2)}
\end{equation}
for various $Q^2$. One can see that -- apart from the region of very large
$x$ -- the dynamical breaking of SU(2) is rather small and could in 
reality well be entirely masked by the explicit breaking in the 
non-perturbative sea input.   
\section{Upper bounds on $A_{TT}$: \label{bounds} Framework}
Now that we have shown that NLO evolution preserves Soffer's inequation, 
we want to utilize it to derive upper bounds on the transverse double spin 
asymmetry to be measured in polarized Drell-Yan muon pair production.
For this purpose we choose again the maximally allowed value 
(\ref{soffer1}) for the transversity distributions, which should yield the 
maximal double spin asymmetry. We employ the same unpolarized and 
longitudinally polarized input distributions as in the previous section, 
along with the same value for the initial scale $Q_0$. 

The scaling variable for the Drell-Yan process is $\tau = M^2/S$, where
$M$ is the invariant mass of the produced muon pair and $\sqrt{S}$ is the
center-of-mass energy of the hadronic collision. Since in unpolarized
reactions only the collision axis is specified, the distribution
of the produced muon pairs cannot depend on the azimuth $\phi$. If the
colliding nucleons are transversely polarized then the collision and 
spin axes specify a plane in space and consequently the polarized
cross section will depend on $\phi$. Instead of working with $\tau$-dependent
cross sections we again prefer Mellin moments defined by
\begin{equation}
\frac{d(\delta)\sigma^n}{d\phi} \equiv \int_0^1 d\tau \tau^{n-1}
\frac{\tau d(\delta)\sigma}{d\tau d\phi} \quad .
\end{equation}
Including NLO corrections to these cross sections one obtains the
generic expression \cite{vowe,coka}
\begin{eqnarray}
\frac{d(\delta)\sigma^n}{d\phi} &=& \frac{\alpha_{em}^2}{9S}
\;(\delta)\Phi(\phi)
\left[ \;(\delta)H^n_q(Q^2_F)\left( 1+ \frac{\alpha_s(Q_R^2)}{2\pi} 
\;(\delta)C_q^{DY,n}(Q_F^2) \right) \right. \label{xsec}
 \nonumber \\
& &
\hspace{2.5cm}+ \left. H^n_g(Q^2_F) \frac{\alpha_s(Q_R^2)}{2\pi}
\;(\delta)C_g^{DY,n}(Q_F^2) \right]\quad ,
\end{eqnarray}
where 
\begin{eqnarray}
(\delta)H_q^n(Q_F^2) &\equiv& \sum_q e_q^2 \left[
(\delta)q_A^n(Q_F^2)\;(\delta)\bar q_B^n(Q_F^2) + 
( A \leftrightarrow B )
\right]\,\,,\label{hq}
\\
H_g^n(Q_F^2) &\equiv& \sum_q e_q^2 \left[
g_A^n(Q_F^2)\left(q_B^n(Q_F^2) + \bar q_B^n(Q_F^2)\right) 
+ ( A \leftrightarrow B ) \right]\quad .\label{hg}
\end{eqnarray}
The dependence on the azimuth is given by $\Phi(\phi)=1$ and 
$\delta \Phi(\phi)=\cos 2\phi$. Integration over $\phi$ thus isolates 
the unpolarized part and $\Phi(\phi)$ is then replaced by $2\pi$.
On the other hand the integration \cite{copi}
$\left(\int_{-\pi/4}^{\pi/4}-\int_{\pi/4}^{3\pi/4}
+\int_{3\pi/4}^{5\pi/4}-\int_{5\pi/4}^{7\pi/4} \right)d\phi$ extracts the
polarized cross section and $\delta \Phi(\phi)$ can then be simply substituted
by 4. In the following we will always assume appropriate integration
over the azimuth.

The unpolarized NLO $\overline{\mbox{MS}}$ QCD coefficients in $\tau$-space 
can be found, e.g., in \cite{fupe}. Their Mellin moments are
\begin{eqnarray}
C_q^{DY,n}(Q_F^2)&=& C_F\left( 4 S_1^2(n) - \frac{4}{n(n+1)}S_1(n) 
+\frac{2}{n^2}+\frac{2}{(n+1)^2}-8 + \frac{4\pi^2}{3} \right ) + 
\nonumber \\
&& 
C_F\left[ \frac{2}{n(n+1)}+3-4S_1(n)\right] \;\ln\left(\frac{M^2}{Q_F^2}
\right)
\quad , 
\\
C_g^{DY,n}(Q_F^2)&=& T_R\left( -2 \frac{n^2+n+2}{n(n+1)(n+2)} S_1(n)
+\frac{n^4+11n^3+22n^2+14n+4}{n^2(n+1)^2(n+2)^2} \right) + \nonumber \\
&& T_R \;\frac{n^2+n+2}{n(n+1)(n+2)}\;\ln\left(\frac{M^2}{Q_F^2}
\right)\quad , 
\end{eqnarray}
where $C_F=4/3$, $T_R=1/2$.
The polarized ones can be found in \cite{voge} and read
\begin{eqnarray}
\delta C_q^{DY,n}(Q_F^2) &=& C_F \left[ 4S_1^2(n) + 12 \left( 
S_3(n)-\zeta(3) \right)+\frac{4}{n(n+1)}-8+\frac{4\pi^2}{3} \right]+ 
\nonumber \\ & &
C_F\left[3-4S_1(n) \right]\;\ln\left(\frac{M^2}{Q_F^2}\right) \quad , 
\\
\delta C_g^{DY,n} (Q_F^2) &=& 0 \label{coefend} \quad .
\end{eqnarray}
In the above formulas we used the abbreviation $S_k(n)=\sum_{j=1}^n j^{-k}$. 
Since there is no gluon transversity distribution for the nucleon, the
gluonic part of (\ref{xsec}) drops out for the polarized case.
The indices $A$ and $B$ in (\ref{hq}) and (\ref{hg}) take into account
the possibility of having two different scattering hadrons, although only $pp$
collisions are planned at the moment.
Finally, $Q_F$ and $Q_R$ in Eqs.~(\ref{xsec})-(\ref{coefend}) 
are the factorization and renormalization scales, respectively,
for which we will choose $Q_F=Q_R=M$ unless stated otherwise.

$Z^0$ production and $\gamma Z^0$-interference can be easily included
by the substitution (see also \cite{coka})
\begin{eqnarray}
e_q^2 &\rightarrow &e_q^2 -8 e_q V_l V_q \kappa 
\frac{M^2(M^2-M_Z^2)}{(M^2-M_Z^2)^2+\Gamma_Z^2M_Z^2}
\nonumber \\
& & +16
(V_l^2+A_l^2)(V_q^2\pm A_q^2)\kappa^2\frac{M^4}{(M^2-M_Z^2)^2+\Gamma_Z^2M_Z^2}
\quad ,
\end{eqnarray}
where
\begin{equation}
\kappa \equiv \frac{\sqrt{2}G_FM_Z^2}{16\pi\alpha_{em}} \quad , \quad 
V_f\equiv T^3_f - 2e_f \sin^2\Theta_W \quad , \quad
A_f\equiv T_f^3\quad .
\end{equation}
The positive sign in front of the $A_q^2$ term is appropriate for the
unpolarized cross section, the negative sign for the polarized one. As usual,
$G_F$ denotes the Fermi constant, $\Theta_W$ the Weinberg angle 
($\sin^2 \Theta_W=0.224$) and $T_f^3$ the third component of the weak isospin.

For examining the perturbative stability of our results we will also
calculate the cross section at LO. In this case one simply needs to set
the QCD coefficients to zero in the above formulas and to replace the 
NLO parton distributions by ones evolved in LO. As LO input distributions 
for (\ref{soffer1}) we will use the unpolarized LO parametrizations of 
\cite{glre2} (neglecting again the SU(2) breaking in the quark sea) and 
the polarized ones of \cite{glre3} at the LO input scale $Q_0^2=0.23\;
\mbox{GeV}^2$.

\section{Results\label{results}}
\begin{figure}
\epsfysize10cm
\leavevmode\epsffile{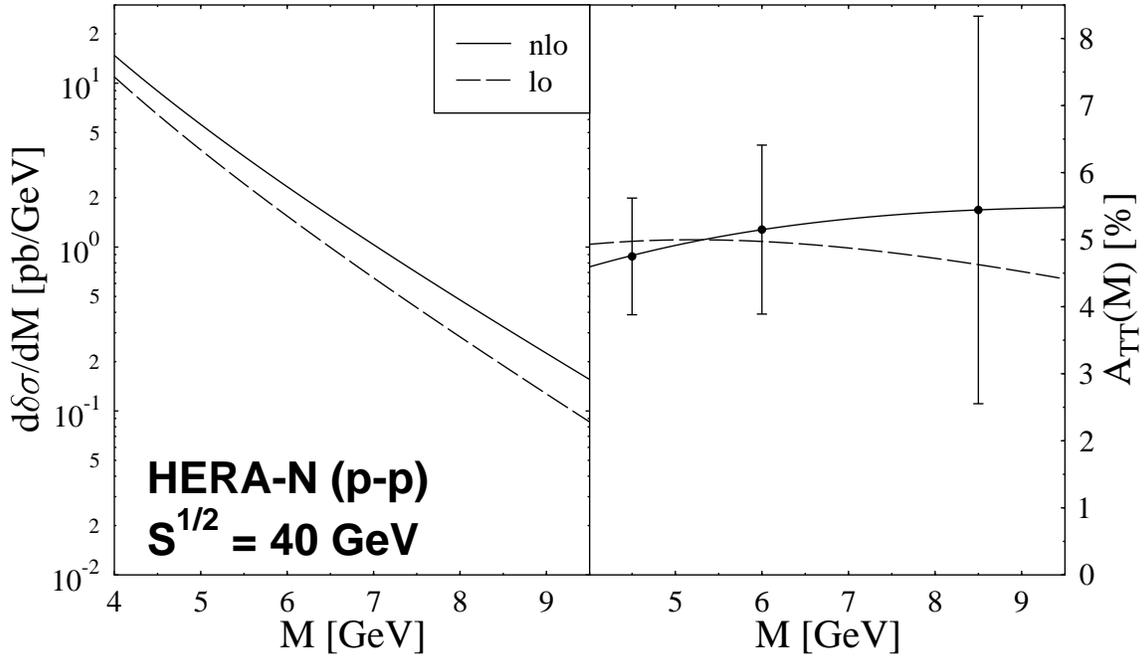}
\caption{\label{fig3}NLO and LO maximal polarized Drell-Yan cross sections 
and asymmetries for HERA-$\vec{\rm N}$. The error bars have been calculated 
according to Eq.~(\ref{aerr}) and are based on ${\cal L}=240\,\mbox{pb}^{-1}$, 
70\% polarisation of beam and target and 100\% detection efficiency.}
\end{figure}

Fig.~\ref{fig3} shows the transversely polarized $pp$ cross section and 
the ``maximal'' double spin asymmetry $A_{TT}$ for $\sqrt{S}=40\;\mbox{GeV}$, 
corresponding to the proposed \cite{nowak} HERA-$\vec{{\rm N}}$ fixed target 
experiment which would utilize the possibly forthcoming polarized 820 GeV 
proton beam at HERA on a transversely polarized target. We show results 
at both LO and NLO. For illustration we have also included the expected 
statistical errors for a measurement of $A_{TT}$ by HERA-$\vec{{\rm N}}$ which
can be estimated from 
\begin{equation}  \label{aerr}
\delta A_{TT} = \frac{1}{P_B P_T \sqrt{{\cal L} \sigma \epsilon}} \; ,
\end{equation}
where $P_B$ and $P_T$ are the beam and target polarizations for which we
will use $P_B=P_T=0.7$. ${\cal L}$ is the  anticipated integrated 
luminosity of ${\cal L}=240\;\mbox{pb}^{-1}$, $\sigma$ the unpolarized
cross section integrated over bins of $M$, and $\epsilon$ the detection
efficiency for which we will take for simplicity $\epsilon=100\%$.
Note that full $4\pi$ coverage of the detector is assumed. Fig.~\ref{fig3} 
shows that the maximal asymmetry for HERA- $\vec{{\rm N}}$ is actually fairly 
large and would be accessible in that experiment. 

\begin{figure}
\epsfysize10cm
\leavevmode\epsffile{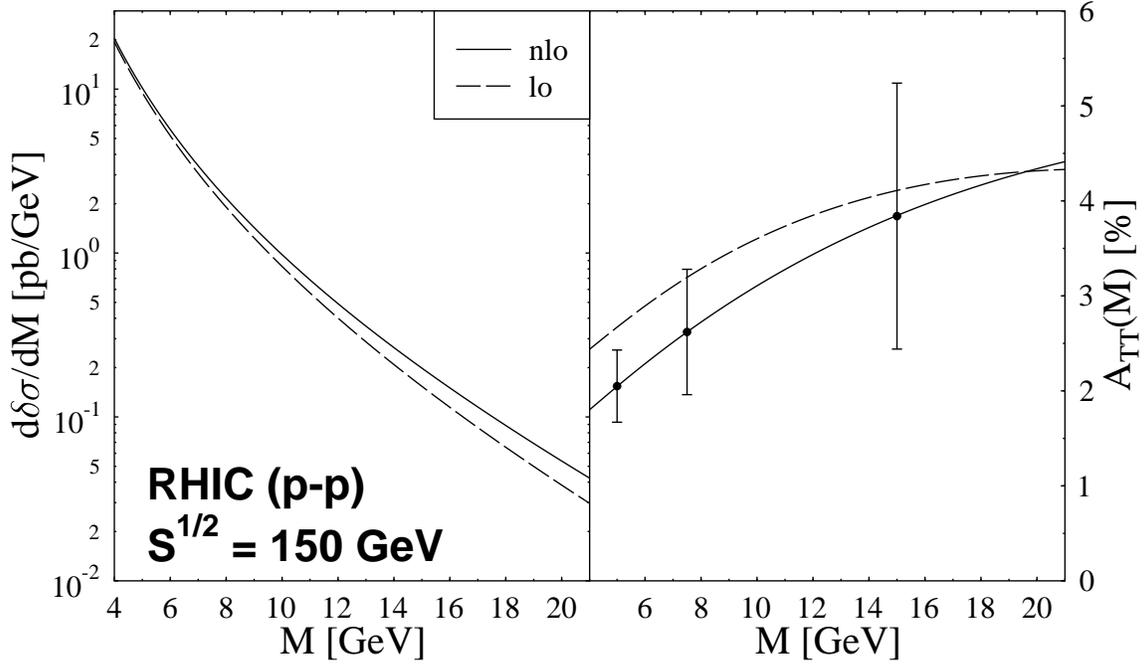}
\caption{\label{fig4}As in Fig.~\ref{fig3}, but for RHIC at $\sqrt{S}=150\,
\mbox{GeV}$ assuming ${\cal L}=240\,\mbox{pb}^{-1}$, 
70\% polarisation of each beam and 100\% detection efficiency.} 

\end{figure}

In Fig.~\ref{fig4} we present results similar to Fig.~\ref{fig3},
but now for $\sqrt{S}=150\;\mbox{GeV}$, corresponding to the RHIC
collider. We note that the region $9$ GeV $\lesssim M \lesssim 11$ GeV
will presumably not be accessible experimentally since it will be 
dominated by muon pairs from bottomonium decays.
Again the predicted maximal asymmetry is of the order of a few 
per cents. From the expected error bars calculated again for $70\%$ beam 
polarization, ${\cal L}=240\;\mbox{pb}^{-1}$ and $\epsilon=100\%$
one concludes that asymmetries of this size should be also well measurable
at RHIC. 

\begin{figure}
\epsfysize10cm
\leavevmode\epsffile{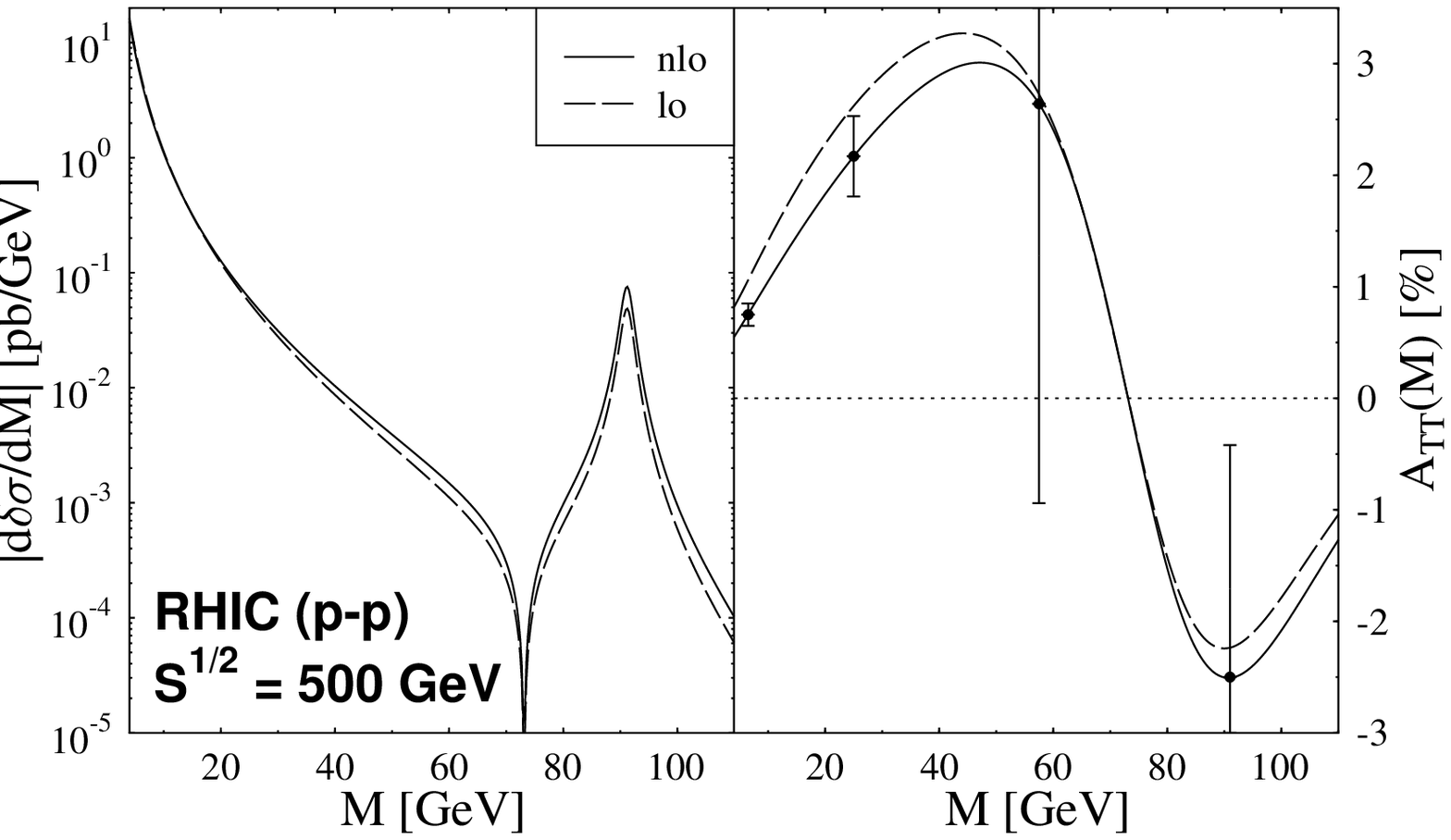}
\caption{\label{fig5}As in Fig.~\ref{fig4}, but for $\sqrt{S}=500\,\mbox{GeV}$
and ${\cal L}=800\,\mbox{pb}^{-1}$.}
\end{figure}

Fig.~\ref{fig5} shows similar results for the high-energy end of RHIC,
$\sqrt{S}=500\;\mbox{GeV}$, where the integrated luminosity is expected 
to be ${\cal L}=800\;\mbox{pb}^{-1}$. It turns out that the asymmetries
become smaller as compared to the lower energies, but thanks to the
higher luminosity the error bars become relatively smaller as well, at least 
in the region $5\;\mbox{GeV}\lesssim M \lesssim 25\;\mbox{GeV}$ where the 
errors are approximately 1/10 of the maximal asymmetry. One can also clearly 
see in Fig.~\ref{fig5} the effect of $Z$-exchange and the $Z$ resonance.

We have already mentioned before that Soffer's inequation does not determine
the sign of $\delta q(x,Q^2)$, so that in principle we have to check all 
different combinations in order to find the ``true'' maximal value for 
$A_{TT}$. It turns out, e.g., that keeping a positive sign only for 
$\delta u_v(x,Q^2)$ leads to a reduction of $|A_{TT}|$ at small 
$M$ but an enhancement at the experimentally not accessible region of 
large $M$. We have checked that for small $M$ the asymmetry takes its 
largest values if all signs are chosen to be positive, as was done in 
Eq.~(\ref{soffer1}) and in the above plots.

A comparison of the LO and NLO results in Figs.~\ref{fig3}-\ref{fig5} answers 
one key question concerning the transversely polarized Drell-Yan 
process: Our predictions for the maximal $A_{TT}$ show good perturbative 
stability, i.e.\ the NLO corrections to the 
cross sections and $A_{TT}$ are of moderate size, albeit not negligible. 
There seems to be a general tendency towards smaller corrections 
when the energy is increasing, which should be mainly due to the 
larger invariant masses probed and to a resulting smaller $\alpha_s (M^2)$.

\begin{figure}
\epsfysize10cm
\leavevmode\epsffile{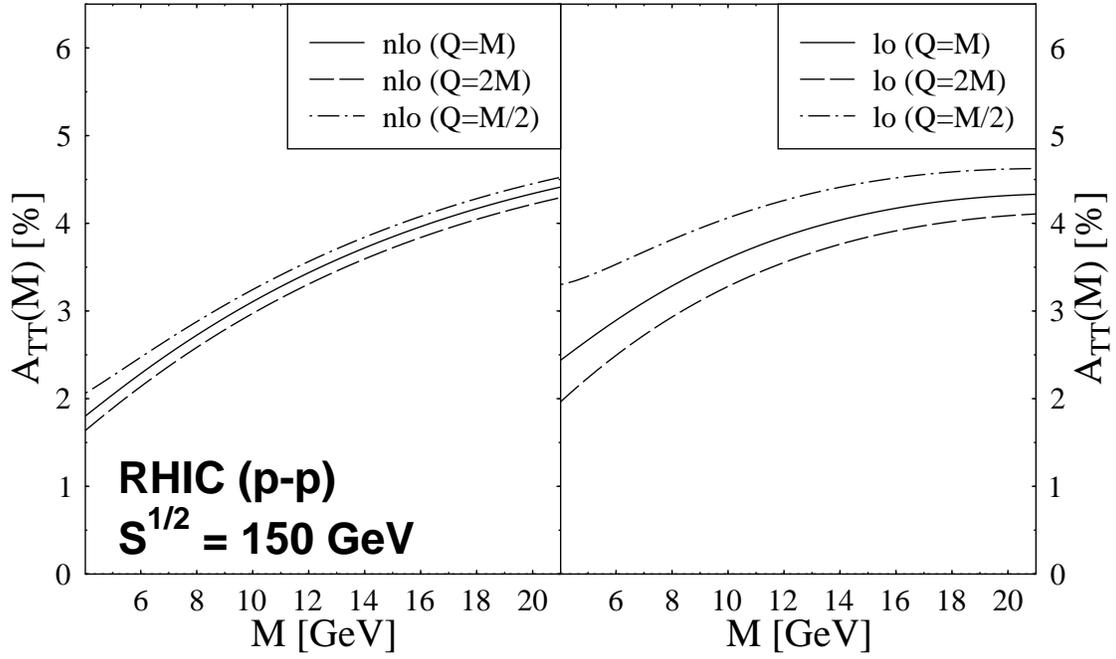}
\caption{\label{fig6}Scale dependence of the LO and NLO asymmetries at
$\sqrt{S}=150\,\mbox{GeV}$. The renormalization and factorization scales 
in (\ref{xsec})-(\ref{hg}) were chosen to be $Q_R=Q_F=M/2,M,2M$. The solid 
line is as in Fig.~\ref{fig4}.}
\end{figure}

Let us finally address some of the the main uncertainties in our predictions 
for the maximal asymmetry $A_{TT}$. The first issue is the scale dependence
of the results. This is examined in Fig.~\ref{fig6} for the case
$\sqrt{S}=150\;\mbox{GeV}$. We plot here the maximal asymmetry in LO and 
NLO, varying the renormalization and factorization scales in the  
range $M/2 \leq Q_F=Q_R \leq 2M$. One can see that already the LO 
asymmetry is fairly stable with respect to scale changes, which is in 
accordance with the finding of generally moderate NLO corrections. The NLO 
asymmetry even shows a significant improvement, so that $A_{TT}$ becomes
largely insensitive to the choice of scale.

\begin{figure}
\epsfysize9cm
\hspace*{3.2cm}
\leavevmode\epsffile{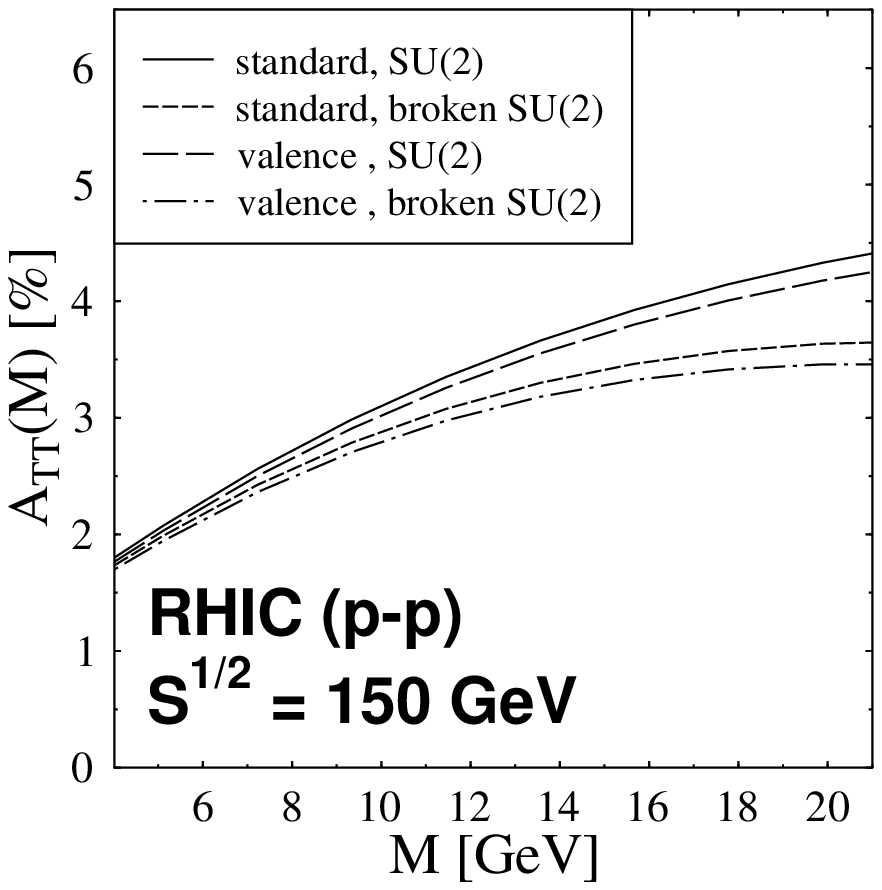}
\caption{\label{fig7}The NLO asymmetry at $\sqrt{S}=150\,\mbox{GeV}$,
using the ``standard'' and the ``valence'' sets of \cite{glre3} for
the $\Delta q(x,Q_0^2)$ in (\ref{soffer1}). Also shown is the change
caused by not neglecting the SU(2) breaking in the $q(x,Q_0^2)$ of 
\cite{glre2} (see text). The solid line is as in Fig.~\ref{fig4}.}
\end{figure}

In order to get a rough idea about the uncertainty caused by our imperfect 
knowledge of the longitudinally polarized parton densities 
$\Delta q(x,Q^2)$, $\Delta g(x,Q^2)$, we have also calculated
the asymmetries using the NLO ``valence'' scenario input distributions of 
\cite{glre3} instead of the ``standard'' ones in (\ref{soffer1}). As can be 
seen in Fig.~\ref{fig7} for the case $\sqrt{S}=150\;\mbox{GeV}$, the  
difference for experimentally significant $M$ turns out to be quite small,
with the predictions based on the ``valence'' scenario distributions
having slightly smaller asymmetries. 

As we already mentioned in Sec.~\ref{evol}, neither the ``standard'' nor 
the ``valence'' scenario parametrizations take into account a possible 
SU(2) breaking in the polarized sea because only neutral current polarized 
DIS data are available at the moment. This led us to neglecting also any
SU(2) breaking in the transversity input densities for our calculations, 
just keeping the dynamical SU(2) breaking produced by NLO evolution
(cf. Fig.~\ref{fig2}). On the other hand, it seems rather likely that
a certain amount of SU(2) breaking -- possibly much more than 
the one generated by evolution -- could be realized in nature. 
One possible way of estimating the uncertainty entering our predictions 
for $A_{TT}$ through this source, is to reintroduce the hitherto neglected 
amount of SU(2) breaking in the unpolarized input densities as fixed 
in the original input distributions of \cite{glre2}. The SU(2) breaking 
will also influence the transversity input via Eq.~(\ref{soffer1}).
The resulting asymmetry is also depicted in Fig.~\ref{fig7}. As can be
seen, the effect is sizeable only at rather large $M$. 
\section{Conclusions \label{conclusions}}
We have shown numerically that Soffer's inequation is preserved by NLO QCD 
evolution provided it is satisfied by the input distributions.

For the first time, we have presented a complete and consistent NLO 
calculation of the transverse double spin asymmetry $A_{TT}$ for the 
Drell-Yan process, employing the NLO corrections to the hard subprocess
cross sections as well as performing the $Q^2$-evolutions in NLO.
Here we have estimated the maximally possible $A_{TT}$ in the framework
of the radiative parton model by assuming that Soffer's inequality is 
saturated at a low hadronic scale. For $\sqrt{S}=40\;\mbox{GeV}$ the maximal 
value of $A_{TT}$ for $pp$ collisions was found to be approximately $4\%$ 
with an expected statistical error for HERA-$\vec{{\rm N}}$ of about $1\%$ at 
an invariant mass of $M=4$ GeV. The situation for RHIC with $\sqrt{S}=150\;
\mbox{GeV}$ turns out to be rather similar. The prospects of measuring 
$A_{TT}$ somewhat improve when going to $\sqrt{S}=500$ GeV where the maximal 
asymmetry is of the order of $1\%$ for small $M$ with an expected relative 
statistical error of approximately 1/10. We emphasize again, however, that our 
results only represent an {\em upper bound} on $A_{TT}$, so that the
``true'' asymmetry may well be much smaller and even experimentally
not measurable. 

Comparing to corresponding LO calculations, we find that the QCD corrections 
turn out to be moderate but non-negligible, putting our predictions on a firm 
basis. We have also examined the main uncertainties of our predictions,
such as the scale dependence of the asymmetry and our imperfect knowledge 
of the longitudinally polarized parton densities to be utilized for the 
saturation of Soffer's inequality at the input scale. We found that 
these uncertainties seem to have rather little impact on our results
in the regions hopefully accessible in future experiments with transversely
polarized protons. 

\noindent {\bf Note added:} After completing this work, we received the paper
\cite{soffernew} in which a mathematical proof of the preservation 
of Soffer's inequality under NLO $Q^2$-evolution is given.
\section*{Acknowledgements}
The work of M.S. was supported in part by the 'Bundesministerium f\"{u}r 
Bildung, Wissenschaft, Forschung und Technologie' (BMBF), Bonn. O.M. and A.S. 
acknowledge financial support from the BMBF and the Deutsche 
Forschungsgemeinschaft.
We thank G.~Bunce for useful information concerning the RHIC luminosities. 
Furthermore, O.M. is grateful to T.~Gehrmann for helpful discussions and 
for providing an evolution program to crosscheck our results.

\newpage

\begin{thebibliography}{99}
\bibitem{raso} J.P.~Ralston and D.E.\ Soper, Nucl. Phys. {\bf B152}, 
                109 (1979).
\bibitem{jaji} R.L.~Jaffe and X.~Ji, Phys. Rev. Lett. {\bf 67}, 552 (1991);
               Nucl. Phys. {\bf B375}, 527 (1992).
\bibitem{arme} X.\ Artru and M.\ Mekhfi, Z. Phys. {\bf C45}, 669 (1990).
\bibitem{rhic} RHIC Spin Collab., D.\ Hill et al., letter of intent 
               RHIC-SPIN-LOI-1991, updated 1993;\\
               G. Bunce et al., Particle World {\bf 3}, 1 (1992); \\
               PHENIX/Spin Collaboration, K.~Imai et al., 
               BNL-PROPOSAL-R5-ADD (1994).
\bibitem{fupe} see, e.g., W.\ Furmanski and R.\ Petronzio, 
               Z. Phys. {\bf C11}, 293 (1982). 
\bibitem{vowe} W.\ Vogelsang and A.\ Weber, Phys. Rev. {\bf D48}, 2073 (1993).
\bibitem{coka} A.P.\ Contogouris, B.\ Kamal and Z.\ Merebashvili,
               Phys. Lett. {\bf B337}, 169 (1994).
\bibitem{voge} W.\ Vogelsang, CERN-TH/97-132, {\tt hep-ph/9706511}.
\bibitem{haka} A.\ Hayashigaki, Y.\ Kanazawa and Y.\ Koike, 
               {\tt hep-ph/9707208},
               to appear in Phys. Rev. {\bf D}.
\bibitem{kumi} S.\ Kumano and M.\ Miyama, Phys. Rev. {\bf D56}, 2504 (1997).
\bibitem{soff} J.\ Soffer, Phys. Rev. Lett. {\bf 74}, 1292 (1995).
\bibitem{baro} V.\ Barone, DFTT 68/96, {\tt hep-ph/9703343}.
\bibitem{masc} X.\ Ji, Phys. Lett. {\bf B284}, 137 (1992); \\
               R.L.\ Jaffe and N.\ Saito, Phys. Lett. {\bf B382}, 165 (1996);\\
               O.\ Martin and A.\ Sch\"afer, Z. Phys. {\bf A358}, 429 (1997);\\
               V.\ Barone, T.\ Calarco, and A.\ Drago, Phys. Rev. {\bf D56},
               527 (1997);\\
               S.\ Scopetta and V.\ Vento, FTUV-29,  
               {\tt hep-ph/9706413}; {\tt hep-ph/9707250}.
\bibitem{glre1} M.\ Gl\"uck, E.\ Reya and A.\ Vogt, Z. Phys. {\bf C48},
               471 (1990).
\bibitem{glre2} M.\ Gl\"uck, E.\ Reya and A.\ Vogt, Z. Phys. {\bf C67},
               433 (1995).
\bibitem{glre3} M.\ Gl\"uck, E.\ Reya, M.\ Stratmann,
                and W.\ Vogelsang, Phys. Rev. {\bf D53}, 4775 (1996).
\bibitem{jama} R.L.\ Jaffe and A.\ Manohar, Phys. Lett. {\bf B223}, 
               218 (1989); \\ 
               X.\ Ji, Phys. Lett. {\bf B289}, 137 (1992).
\bibitem{ross} D.A.\ Ross and C.T.\ Sachrajda, Nucl. Phys. {\bf B149},
497 (1979).
\bibitem{svw} M.\ Stratmann, W.\ Vogelsang, and A.\ Weber, 
Phys. Rev. {\bf D53}, 138 (1996).
\bibitem{copi} J.L.\ Cortes, B.\ Pire, and J.P.\ Ralston,
               Z. Phys. {\bf C55}, 409 (1992).
\bibitem{nowak} V.A. Korotkov and W.-D. Nowak, DESY-97-004,
                {\tt hep-ph/9701371}, talk presented at the 2nd ELFE
                workshop, St.\ Malo, France, Sept.\ 1996.
\bibitem{soffernew} C.\ Bourrely, J.\ Soffer and O.V.\ Teryaev, CPT-97-P-3538, 
                {\tt hep-ph/9710224}.
\end{thebibliography}
\end{document}